\documentclass[sigconf]{acmart}              

\AtBeginDocument{%
  \providecommand\BibTeX{{%
    \normalfont B\kern-0.5em{\scshape i\kern-0.25em b}\kern-0.8em\TeX}}}

\usepackage{graphicx} 

\usepackage{color}
\usepackage{multirow}
\usepackage{amsmath}
\usepackage{algorithm}
\usepackage[noend]{algpseudocode}
\usepackage{mathtools}
\usepackage{svg}
\usepackage{amsmath}
\usepackage[labelfont=bf]{caption}

\begin{document}

\title{Efficient On-Chip Communication for Parallel Graph-Analytics on Spatial Architectures}

\author{Khushal Sethi}
\email{khushal@stanford.edu}
\affiliation{%
  \institution{Stanford University}
}

\begin{abstract}
Large-scale graph processing has drawn great attention in recent years. Most of the modern-day datacenter workloads can be represented in the form of Graph Processing such as MapReduce etc. Consequently, a lot of designs for Domain-Specific Accelerators have been proposed for Graph Processing.
Spatial Architectures have been promising in the execution of Graph Processing, where the graph is partitioned on several nodes and each node works in parallel. \\
We conduct experiments to analyze the on-chip movement of data in graph processing on a Spatial Architecture. Based on the observations, we identify a data movement bottleneck, in the execution of such highly-parallel processing accelerators. To mitigate the bottleneck we propose a novel power-law aware Graph Partitioning and Data Mapping scheme to reduce the communication latency by minimizing the hop counts on a scalable network-on-chip. The experimental results on popular graph algorithms show that our implementation makes the execution $ 2-5 \times$ faster and $2.7-4 \times$ energy-efficient by reducing the data movement time in comparison to a baseline implementation. 

\end{abstract}

\ccsdesc[500]{Computer systems organization~Spatial Architecture}
\ccsdesc[500]{Networks~Network on chip}
\ccsdesc[500]{Computer systems organization~Domain Specific Accelerators}
%
\keywords{Graph processing, Application-Specific Accelerators, On-chip Communication, Spatial Architecture, Content-Addressable Memory}

\maketitle
\section{INTRODUCTION}
Graphs are widely used to model both data and relationships in real-world problems, such as knowledge mining, modelling complex systems, network analytics etc. 

Popular graph algorithms such as BFS, SSSP, PageRank are widely used in real world server workloads. The demand of graph processsing workloads is increasing rapidly with the increase in internet usage.  

Hence, a large number of systems for efficient execution of large graph workloads on CPU, GPUs, FPGAs, Hybrid Memory Cube (HMC) and also custom accelerators have been proposed in the past \cite{gui2019survey}.

In this paper, we analyze a Spatial architecture based domain-specific accelerator that utilizes Content Addressable Memories to accelerate Graph Applications efficiently with high storage density and efficiency by modeling both computation and memory in graph processing. Although, several works for graph processing in-memory have been investigated, we recognize a data movement bottleneck, in the highly-parallel execution of such accelerators. To mitigate the bottleneck, we propose a novel "power-law" aware Graph Partitioning and Data Mapping scheme to reduce the communication latency. Our mapping scheme minimizes hop counts on a scalable Network on Chip. 
The experimental results on popular graph algorithms, show that our implementation achieves $2-5 \times$ speedup and $2.7-4 \times$ energy reduction in comparison to the baselines. \\
The paper is organized as follows : 
Section 2 provides a background on the Graph Frameworks (Vertex-Centric Processing) and Spatial Architectures used in this paper.
Section 3 describes our Experimental Setup, Section 4 contains the discussion for analyzing data movement and our proposed data mapping scheme. Section 5 lists our key results and performance of our implementation. Section 6 and 7 contain previous related work and conclusions respectively. \cite{bashir2019power, sethi2021efficient, ji2020reconfigurable,ji2020compacc, sethi2020nv, sethi2020design, sethi2018low, sethi2019optimized, sethi2022dragon}

\section{Background}

\begin{table*}[!htbp]
  \centering
 \caption{\textbf{Vertex-centric Programming Model}}
    \begin{tabular}{|c|c|c|c|c|}
    \hline \textbf{Application} & Property & Process & Reduce & Apply \\
     \hline \textbf{Breadth First Search} & Depth & eProp=u.Prop+1 & u.Temp=min(e.Prop,u.Temp) & u.Prop=min(u.Prop,u.Temp)  \\
  \hline \textbf{SSSP} & Distance & eProp=u.Prop+weight & u.temp=min(eProp, u.Temp)& u.Temp=min(u.Prop,u.Temp) \\
    \hline \textbf{Page Rank} & Page rank score & eProp=u.prop+weight & u.Temp=u.Temp+eProp & u.Prop=a*u.Prop+base \\
    \hline
    \end{tabular}%
  \label{tab:parameter}%
\end{table*}%

\subsection{Graph Frameworks}

 \begin{algorithm}
\caption{Vertex-Centric Model}
\label{euclid}
\begin{algorithmic}
\State \emph{\textbf{Process-Phase}}:
\For {$u : ActiveList$}
    \For {$v : OutNeighbour(u)$}
        \State $eProp (u,v)= process(v.Prop, edge(v,u))$ 
    \EndFor {}
\EndFor {}
\State \emph{\textbf{Reduce-Phase}}:
\For {$u : AllVertices$}
\For {$v : InNeighbour(u)$}
\State $eProp(u,v)= reduce(u.Temp, edge(v,u))$ 
\EndFor
\EndFor
\State \emph{\textbf{Apply-Phase}}:
\For {v: $AllVertices$}
\State $v.Prop = Apply(v.Temp, v.Prop))$ 
\EndFor
\end{algorithmic}
\end{algorithm}

The vertex-centric programming model for graph processing consists of three phases : Process, Reduce and Apply. This functioning takes several iterations for processing to converge, in each iteration the search frontier of vertices increases. Algorithm 1 describes the vertex-centric programming model in graph-processing. The execution is stopped when a converged solution is found. The Process-Reduce-Apply functions for processing graphs for popular algorithms in shown in Table 1. The process and apply phases do not contain dependencies and can be executed in parallel where-as reduce phase should be executed serially, but can be parallelized by addition of extra fields in the Graph \cite{zhou2019gram}.


\begin{figure}[!thpb]
    \centering
    \includegraphics[scale=0.45]{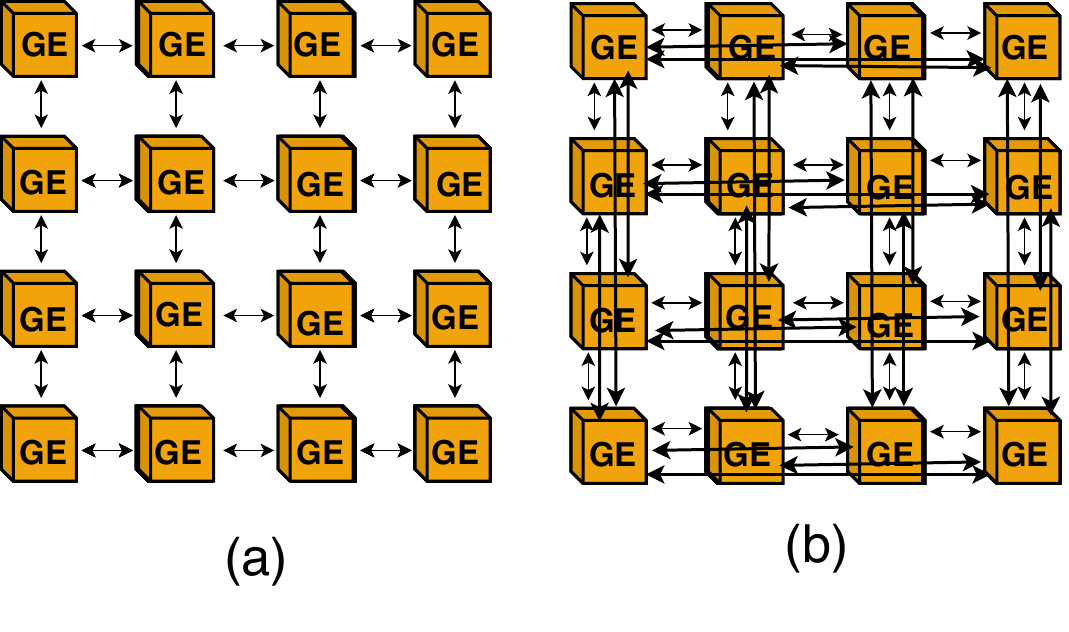}
    \caption{Spatial Architecture for Graph Processing. Each Processing Engine is connected with a NOC (a) 2D Mesh (b) Flattened Butterfly.}
    
    \label{fig:my_label}
\end{figure}

\begin{figure}[!thpb]
    \centering
    \includegraphics[scale=0.55]{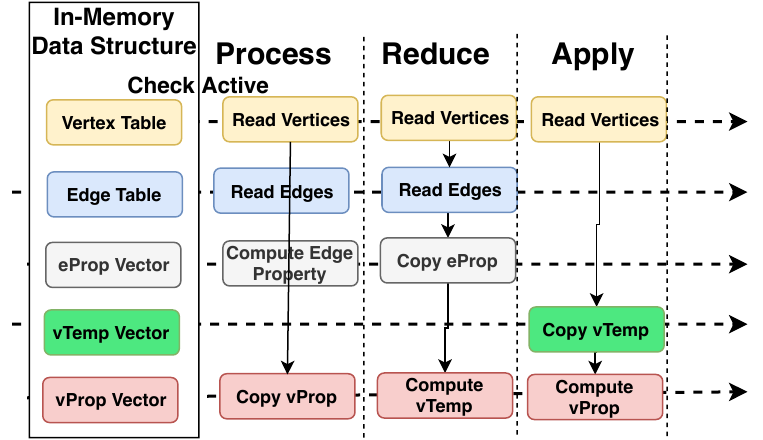}
    \caption{In-Memory Graph Processing Flow}
    \label{fig:my_label}
\end{figure}

\subsection{Spatial Architectures based Domain Specific Accelerators}
Many spatial graph processing architectures/accelerators based have been proposed in the literature (e.g., ReRAM based RPBFS \cite{han2018novel}, GraphR \cite{song2018graphr}, GraphSAR \cite{dai2019graphsar} and HyVE \cite{huang2018hyve}), achieving great speedup and energy efficiency improvement. The implementations in GraphR, GraphSAR and HyVE utilize the matrix-vector multiplication capabilitites of ReRAM in analog domain. In the design of RPBFS, GraphR, and HyVE, edges in the graph are stored using the edge list format. GraphR converts stores the graph in edge list format and converts it into adjacency matrix for MVM operation and stores back into the ReRAM crossbar. GraphSAR proposed a sparsity-aware data partitioning for improving the performance of GraphR. HyVE proposed a design for a memory hierarchy of hetergenous storage over various levels of caches and ReRAM to optimize its execution, which involves off-chip data movement. HyVE uses conventional CMOS circuits to process the subgraphs edge by edge. 
GRAM \cite{zhou2019gram} uses 2-D ReCAM based architecture, which operates search operations in the digital domain and all the blocks are serially connected to a memory controller which dictates the execution and data movement on a shared bus. 
RADAR\cite{huangfu2018radar} similar to us analyzes 3D-ReCAM for genome sequencing.\\
The high-level architecture of a spatial architecture is shown in Fig. 1. It typically consists of multiple Graph Engines, the fundamental computational blocks.  Each Graph Engine can execute in parallel which provides a high degree of parallelism for execution in different graph processing stages. The On-Chip Controller is responsible for broadcasting operations to each Graph Engine. \\
Each Graph Engine may consist of input and output buffers, an on-node storage (for storing the partitioned Graph), Arithmetic Logic Units (ALUs) for post processing the data. \\
The Graph Engines are connected in a "memory-centric" NOC (such as 2D Mesh, DragonFly, Torus) structure by which they can route packets of data. 
\subsection{Data Flow}
The processing flow in Fig 2. follows through an in-memory layout. The graph structure in memory is represented by a Edge Table ET and a Vertex Table VT. The edge table and vertex table are arranged column-wise in the dense memory to provide fast content addressable search during the execution \cite{zhou2019gram}. For graph G, with M edges and N vertices, properties related to Edges and Vertices have to be stored in the memory. 
The Process-Reduce-Apply execution involves movement of data between these Graph Engines shown in Fig. 1. A parallel search in the Edge list looks up the data and neighbours. Thus large data movement happens between Edge List and Vertex Prop (vprop), Vertex Temp (vtemp) data structures. No movement of data is between the Edge Property (eprop) and Edge list during the execution. Similarly, data movement happens between the Edge Property (eprop) and Vertex Prop (vprop), Vertex Temp (vtemp) data structures. The data-structures will be distributed on the spatial architecture, on-chip content addressable memory at each Graph Engine containing only a portion of the data structures.
\begin{table}[!htbp]
  \centering
 \caption{\textbf{Graph Processing Workloads}}
    \begin{tabular}{|c|c|c|c|}
    \hline Graph & $\#$Vertices & $\#$Edges & Description \\
    \hline amazon & 304K & 4.3M & Purchasing Network\\
    \hline soc-pokec & 1.6M & 30.6M & Social Network \\
    \hline wiki-topcats (wiki) & 1.8M & 28.5M & Hyperlinks of Wikepedia \\
    \hline ljournal & 5.4M & 78M & Live Journal \\
    \hline 
    \end{tabular}%
  \label{tab:parameter}%
\end{table}%
\section{Experimental Setup}
We study the design in a trace-driven simulator to model the functionality for the spatial architecture considered. We modified open-source GraphMAT \cite{sundaram2015graphmat} to generate traces for our simulation.
We demonstrate on three commonly used graph-processing algorithms: Breadth First Search (BFS), Single Shortest Source Path (SSSP), and Page-Rank(PR). The vertex-centric model for each is shown in Table 1. The experiments are run on real world graphs obtained from the SNAP Datasets \cite{snap}.
\section{Analysis of Data Movement}
\begin{figure}[!thpb]
    \centering
    \includegraphics[scale=0.40]{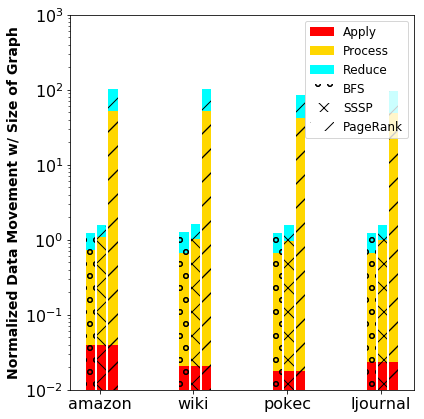}
    \caption{Total On-Chip Data Movement (Normalized with the size of Graph) in Graph Processing of different algorithms for Process and Reduce phase is shown. }
    \label{fig:data}
\end{figure}

The amount of data movement (normalized w/ size of graph) is shown in Fig. \ref{fig:data} for 3 different algorithms considered in our experiments. The amount of data transferred is nearly equal for process and reduce phase, whereas data movement in apply phase is negligible compared to the first two phases. PageRank requires more data-movement because it takes more number of iterations to converge, hence data is updated several times. 
The bulk of data movement in Process phase occurs from reading the Edge Table and correspondingly transferring data to Vertex Property (vprop) and then to Edge Property (eprop) for updates (Figure 1). In reduce phase also, the data movement between arises from updating Temporarily stored vertex data (vTemp) by Edge Property (eprop) data structure and reading Edge Table (ET) for neighbours (Figure 1). 
If additional book-keeping is done for parallel-reduce operations (as suggested in \cite{zhou2019gram}) the network traffic in considerably increased further. \\
Figure 4 shows the skewed distribution of edges on different vertices, where sometimes even less than 10\% of vertices are connected in 90\% of the edges. This skewness is even greater in larger graph-databases. This skewness is explained by power-law which is that the degree distribution of vertices vector n(d) with non-zero entries that follows the relation
\begin{equation}
    n(d) \propto 1/d^{\alpha}
\end{equation}
where $\alpha > 0$ is the slope of the power law (in log scale), d is the number of edges in the vertex of a graph, and n(d) is the number vertices with a specific degree d.
\section{Data Mapping Optimization}
The NoC topology can be represented as a directed graph P(U, F) with each vertex $u_i \in U$ representing a node in the topology and the directed edge $f_{i,j} \in F$ representing a physical link between the vertices $u_i$ and $u_j$ . If the weights of all connections  are considered equal ($f_{i,j} \in \{0,1\}$) then, latency (T) of a packet through a network can be summarized as \begin{equation}
T = H*(T_r+T_w)  \end{equation} where H is the hop count and $T_r$ is the per-hop latency through the switch and $T_w$ is the channel link latency. For minimizing the average hop count an efficient data mapping is required. \\
The in-memory graph structure denoted by M (V,E) consists of Edge Table (ET) where edges are denoted by E, Vertex Temp (vtemp), Vertex Property (vprop) and Edge Property (eprop) data structures, with data movements between them as described in the previous sections. \\
The "power-law" aware graph partitioning and mapping in each of the data structures and their spatial arrangement is described in Algorithm 2. 

\subsection{Graph Partitioning to Nodes :} The graph is partitioned in a source-cut (edge) fashion. As a first step, the vertices are sorted by descending order of degree, for ease. The higher degree vertices will have more number of edges (combined) compared to lower degree vertices. \\
For edge-list partitioning, each edge stored in a Node will have one vertex from a small set of higher degree vertices, from the top of sorted vertex list. In other words, the edges from these higher degree vertices are distributed on to the nodes.

Now, the edges/vertices are allocated to nodes in a load-balancing aware scheme.

Load balancing is carried by using modulo scheduling, which means among the $v$ vertices in sorted vertex list, are distributed in a cyclic fashion to the nodes.

 \begin{algorithm}
\caption{Graph Partitioning in Nodes}
\label{euclid}
\begin{algorithmic}[1]
\State {Inputs : NoC Topology Graph P (U,F) and M (V,E)}
\State {Output : M (V,E) $\to$ P (U,F) : $V \to U$, such that, $v_{i} \in V$,  $u_{j} \in U$ and map ($v_{i}$) = $u_j$.} 
\State {V' $\gets$ sort all v $\in V$ in order of out-degree, V $\gets$ V'}
\State \emph{\textbf{Edge-list Partitioning $(\forall u \in U, u.index \in \{1,4\})$}}:
\For {$v \in V\%{num\_of\_nodes} \ and \ e \ni e.src=v$}
\While {$u.size < u.maxsize$}
 \State $u \gets e$
\EndWhile
\State $u.rank=min(v) \forall v \in u$ 
\EndFor
\State \emph{\textbf{Vertex Partitioning $(\forall u \in U, u.index \in {2,3})$}}
\For {$v \in V\%{num\_of\_nodes}$}
\While {$u.size < u.maxsize$}
\State $u \gets v$
\EndWhile
\State $u.rank=min(v) \forall v \in u$
\EndFor
\end{algorithmic}
\end{algorithm}

\subsection{Scalable Placement of Nodes :}
Since each graph engine is an NOC node, each NOC node $u_{i} \in U$ is assigned a coordinate $(x_{i},y_{i})$ and an additional fields $rank_{i}$ and $index_{i}$, where the index is integer-allocated (1-4) based on the data structure (Edge Table, Vertex Prop, Vertex Temp, Edge Prop) in the following order, and a rank is allocated to link corresponding vertex-cut Nodes with different index.

Suppose a node $u_{i}$ contains data for Vertex Property, then the $rank_{i}$ indicates the minimum vertex id contained and $index_{i}=0$. 

Then with the corresponding node $u_{j}$ containing Edge Table with same $rank$, we have $f_{i,j}  = 1$. 

\begin{figure}[!thpb]
    \centering
    \includegraphics[scale=0.45]{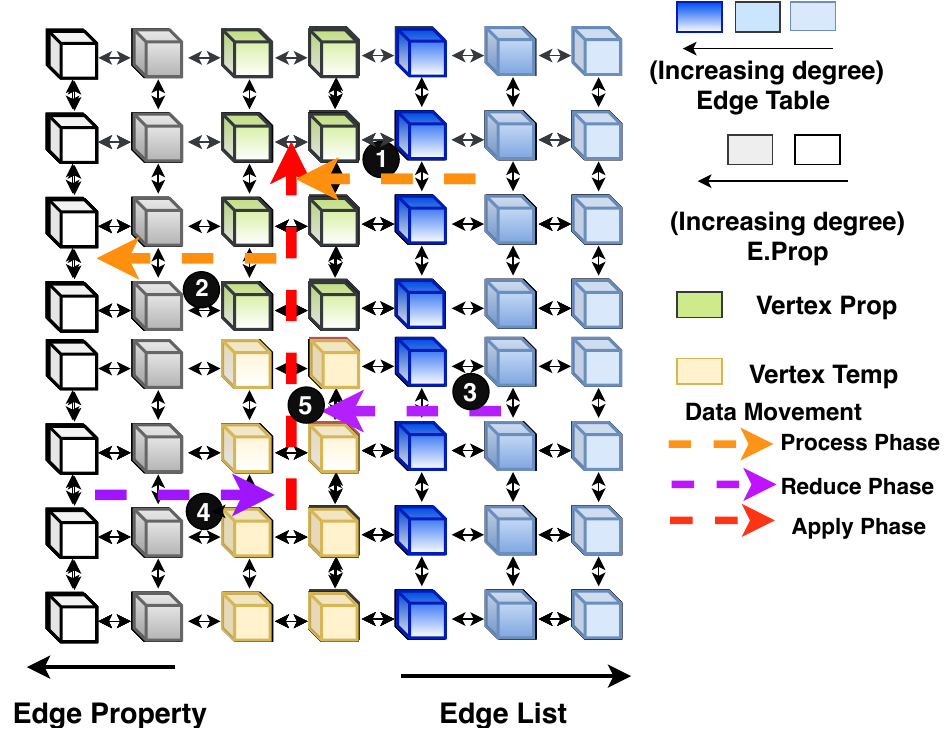}
    \caption{Schematic of the Placement Strategy of In-Memory Data Structures for Graph Processing.}
    \label{fig:map}
\end{figure}

The initial arrangement of In-Memory data structures is shown in Fig \ref{fig:map}, where the mapping starts from the center of the cluster. 

We define regularity constraints on each on the data-structure, for predictable and homogeneous data-transfers. 

For this, different data-structures are placed in columnar and row fashion in the spatial architecture, such that positions of the data-structure between which data-transfer is higher should be closer. The formal constraints for such are defined in Algorithm 3.

\begin{algorithm}
\caption{Scalable Placement of Nodes}
\label{euclid}
\begin{algorithmic}[1]
\State {Inputs : NoC Topology Graph P (U,F) and M (V,E)}
\State \emph{\textbf{Populating Topology Graph Edges}}
\State $\forall u_i,u_j \in U \ni u_{i}.rank=u_j.rank $ and $u_{i}.index \in \{1,4\}, u_{j}.index \in \{2,3\}, \ do \ f_{i,j}=1$
\State \emph{\textbf{Constraints for Regular-Scalable Structure}}
\State $\forall u \in U, u.index=1, u.y>0$
\State $\forall u \in U, u.index=4, u.y<k$
\State $\forall u \in U, u.index=2, 0<u.y<k, u.x>0$
\State $\forall u \in U, u.index=3, 0<u.y<k, u.x>0$
\end{algorithmic}
\end{algorithm}

\subsection{ILP Formulation for Optimal Placement :}

For the determmining the 2-D coordinates of each node, on the network on chip mesh, we formulate it as an Integer Linear Programming (ILP) problem that finds the coordinates such that it minimizes the total cost (hop count) of the data transfer between nodes. 

As defined above, the data-transfer happens between the nodes $u_i$ and $u_j$ for which the edge $f_{i,j}  = 1$. The cost of data-transfer between these nodes is equivalent to the topological distance between them. In case of a 2-D mesh this topological distance would just be the $L_1$ norm distance.

This optimization (in Algorithm 4) is solved computationally with the constraints defined in Algorithm 3 as an ILP. 

 \begin{algorithm}
\caption{ILP Formulation of the Optimization Objective}
\label{euclid}
\begin{algorithmic}[1]
\State {Inputs : NoC Topology Graph P (U,F)}
\State {Output : $(x_i,y_i) \leftrightarrow u_i$} 
\State H = min $\sum_{i} \sum_{j} f_{ij} cost((x_i,y_i) \leftrightarrow (x_j,y_j))$
\State Where $cost((x_i,y_i) \leftrightarrow (x_j,y_j))$ depends on the Topological Distance,
\State For 2-D Mesh : $cost((x_i,y_i) \leftrightarrow (x_j,y_j))$ = $(\mid x_i -x_j \mid+ \mid y_i - y_j \mid)$
\State For Flattened Butterfly : $cost((x_i,y_i) \leftrightarrow (x_j,y_j))$ = $(\mid x_i -x_j \mid+ \mid y_i - y_j \mid)$
\end{algorithmic}
\end{algorithm}

\begin{figure}[!thpb]
    \centering
    \includegraphics[scale=0.45]{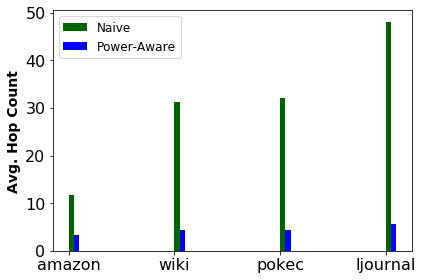}
    \caption{Decrease in the Average Hop Count with our proposed placement strategy (Assuming 2D Mesh Topology).}
    \label{fig:hops}
\end{figure}

Fig. \ref{fig:hops} shows the decrease in hop-count from the proposed strategy after finding the node coordinates by solving the ILP, and using those coordinates in our simulation in a comparison with the randomized mapping strategy for the nodes.

\section{Results}
\subsection{Overall Performance}
\begin{figure}[!thpb]
    \centering
    \includegraphics[scale=0.3]{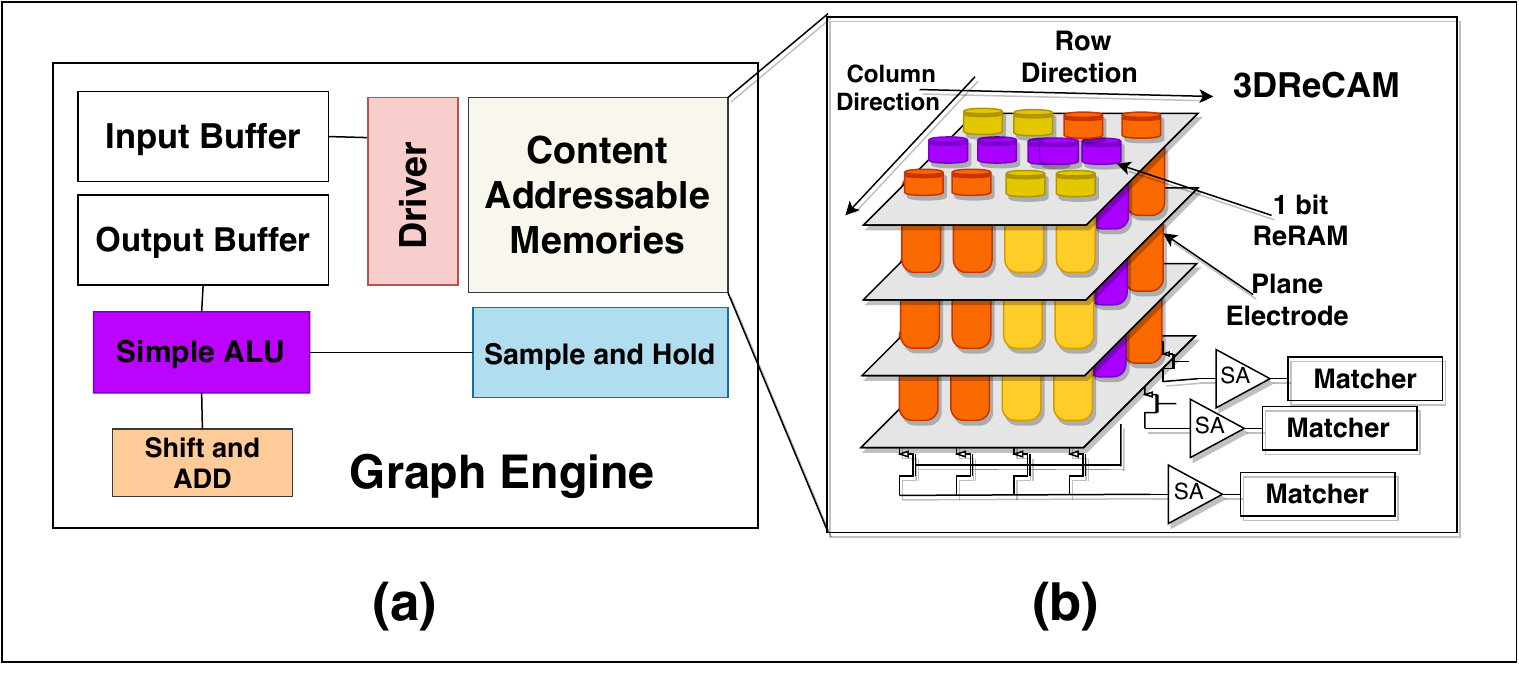}
    \caption{(a) Node Configuration as presented in GRAM \cite{zhou2019gram} showing I/O Buffers, ALUs and Content Addressable Memories (b) 3DReCAM implementation }
    \label{fig:arch}
\end{figure}

\begin{figure*}[!thpb]
    \centering
    \includegraphics[scale=0.5]{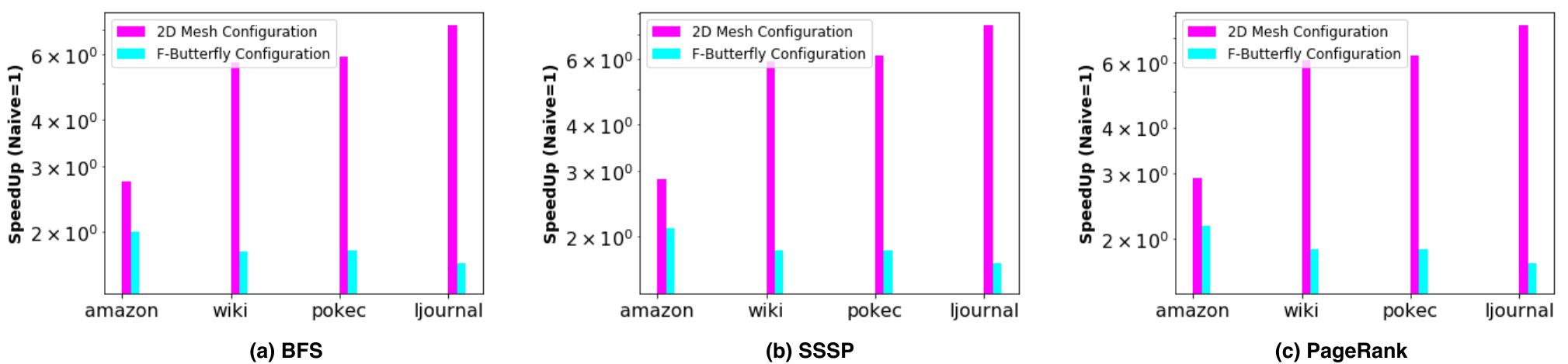}
    \caption{Speedup from our proposed optimization due to lower data movement : for a 3D-ReCAM (Resistive Content Addressable Memory) Architecture with 2D-Mesh and F-Butterfly NoC Configuration}
    \label{fig:result}
\end{figure*}

\begin{figure*}[!thpb]
    \centering
    \includegraphics[scale=0.5]{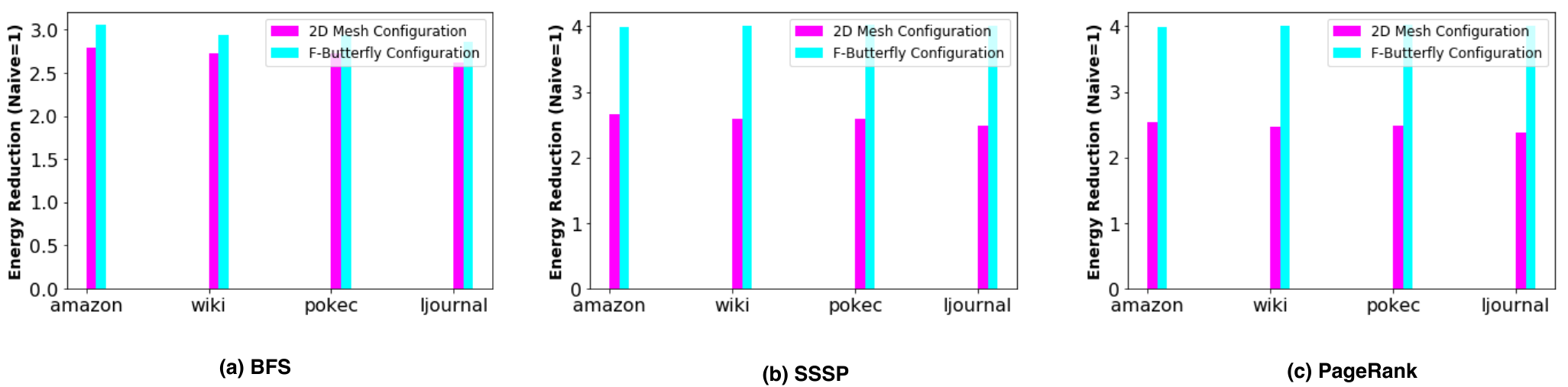}
    \caption{Energy Consumption Reduction : for a 3D-ReCAM (Resistive Content Addressable Memory) Architecture with 2D-Mesh and F-Butterfly NoC Configuration}
    \label{fig:result}
\end{figure*}

To characterize the benefit of our optimization compared to previous works, we consider the spatial architecture (100 Mhz) configuration as described in \cite{zhou2019gram} with a 3D ReCAM in Graph Nodes for doing Parallel Search (shown in Fig ). We analyze the effect of utilizing 2D-Mesh and F-Butterfly Network on Chip configuration in the Spatial Accelerator.

The parameters for the simulation are taken from NVSim-CAM \cite{li2016nvsim} and Destiny \cite{poremba2015destiny} modelling tools. The parameters of the 2-D Mesh NOC used are configured from ORION and input-output buffers are modelled using CACTI 6.5 \cite{muralimanohar2009cacti}. These are listed in Table \ref{tab:parameter}.

We use the metric of performance in terms of speed-up in execution time and energy consumption, where our proposed optimization ((in 2D Mesh NoC Configuration) results in a on average (geometric mean) $5 \times$, $5.2 \times$ and $5.4 \times$ faster and $2.7 \times$, $2.58 \times $ and $2.4 \times$ energy-efficient for BFS, SSSP and PR respectively. 

Similarly, for the flattened butterfly configuration, the execution becomes $1.8 \times$, $1.85 \times$ and $1.9 \times$ faster and $2.9 \times$, $ 4.0 \times $ and $ 4.0 \times$ energy-efficient (in 2D Mesh NoC Configuration) for BFS, SSSP and PR respectively.

\begin{table}[!htbp]
  \centering
 \caption{\textbf{Parameters Used}}
    \begin{tabular}{|c|c|}
    \hline \textbf{CAM Parameters (\cite{zhou2019gram})} & \\

     \hline MAT Size & 1024*1024*8 \\	
     \hline Per Engine Capacity & 1MB \\
     \hline Word & 64 bits \\
    \hline \textbf{NOC Parameters (\cite{zhou2019gram,kahng2011orion})} & \\
    \hline Frequency & 1GHz (1ns)	\\
    \hline Packet Size & 8 bytes	\\
    \hline Latency of hops & 1ns\\
    \hline Ports & 4 \\
    \hline Configuration & 2D Mesh \\
    \hline 
    \end{tabular}%
  \label{tab:parameter}%
\end{table}%

\section{Related Work}
GraphP proposed a HMC-based graph processing system that reduces communication, by two techniques : “Source-cut” partitioning (which utilizes duplicating data), and Hierarchical "non-blocking" communication and for efficiently routing traffic. We have proposed another considerable optimization for reducing communication based on efficient mapping of data NOCs by utilizing "power-law". Our work is hence, orthogonal to their approach and can be improved by combining their approaches.

Previous works have explored mapping static and dynamic tasks efficiently to processor-centric NOCs (\cite{sahu2013survey,yang2016application,maqsood2015dynamic,sahu2014application,wu2015efficient}) by using Integer Linear Programming, Heuristic Search and Deterministic Search. However, this has not been explored for memory-centric PIM systems with high parallelism which have a high programmable complexity. In such systems a regular structure is also required in mapping the in-memory data structures to the nodes which is scalable and easy to implement in the memory controller. This requires good choice of methodologies for efficient and regular data and processing mapping to network-on-chip for processing-in-memory technologies. 
\section{Conclusions}
In this work we analyze spatial architectures proposed to accelerate Graph Applications. 

Content Addressable Memories allow faster search, allowing faster execution, but in the fast execution, the on-chip traffic becomes a bottleneck in the execution time.

By utilizing the inherent characteristics of such Graph Processing Workloads, an efficient mapping is designed to reduce the on-chip traffic and minimize the communication latency between the spatial blocks. The experimental results show that our method can  minimize the hop counts for routing packets of data and also significantly decrease the execution time and energy consumption on CAM-based Spatial Architectures.

\bibliographystyle{IEEEtran}
\bibliography{ref}
\end{document}